\documentclass[10pt,twocolumn]{article}
\usepackage[dvips]{graphicx}
\usepackage{subfigure}
\usepackage{amsmath}
\usepackage{color}
\usepackage{sidecap}
\usepackage{lipsum}
\usepackage{widetext}
\usepackage[font={small}]{caption}

\usepackage[numbers,sort&compress,comma]{natbib}
\usepackage{float}
\usepackage{setspace}
\usepackage{multirow}
\usepackage{makecell}
\usepackage[margin=0.5in]{geometry}

\newcommand{\comments}[1]{} 

\renewcommand{\vec}[1]{{\mathbf{#1}}}

\date{}
\title{\Large{\textbf{
Reducing the effect of Metropolization on mixing times in molecular dynamics simulations
}}}
\author{Jason A. Wagoner\footnote{Department of Chemistry, Stanford University}
 and
Vijay S. Pande \footnotemark[1] \footnote{Departments of Structural Biology and Computer Science, Stanford University}
 \thanks{To whom correspondence should be addressed:  pande@stanford.edu}}

\begin{document}

\maketitle

\textbf{Molecular dynamics algorithms are subject to some amount of error dependent on the size of the time step that is used.  This error can be corrected by periodically updating the system with a Metropolis criteria, where the integration step is treated as a selection probability for candidate state generation.  
  Such a method, closely related to generalized hybrid Monte Carlo (GHMC), satisfies the balance condition by imposing a reversal of momenta upon candidate rejection.  In the present study, we demonstrate that such momentum reversals can have a significant impact on molecular kinetics and extend the time required for system decorrelation, resulting in an order of magnitude increase in the integrated autocorrelation times of molecular variables for the worst cases.  We present a simple method, referred to as reduced-flipping GHMC, that uses the information of the previous, current, and candidate states to reduce the probability of momentum flipping following candidate rejection  while rigorously satisfying the balance condition.  This method is a simple modification to traditional, automatic-flipping, GHMC methods and significantly mitigates the impact of such algorithms on molecular kinetics and simulation mixing times.
} 

\section{Introduction}

Molecular dynamics integration algorithms can be subject to some amount of error resulting from finite time step effects, resulting in sampling that deviates from the desired stationary distribution.  These effects can be corrected using hybrid Monte Carlo (HMC) \cite{Duane1987} or generalized hybrid Monte Carlo (GHMC) \cite{Horowitz1991} algorithms that implement a Metropolis acceptance criteria in which the dynamics paths are treated as candidate selection probabilities. 
 If the momenta are not randomized at the refreshment steps (as is done in the traditional HMC algorithm) the detailed balance condition for this method imposes that the algorithm either reverse momenta upon rejection or that it reverse momenta upon acceptance (indicative of the GHMC algorithm) \cite{Horowitz1991,Kennedy1996}.
 
\begin{figure}
 \centering
\includegraphics[width=8.0 cm]{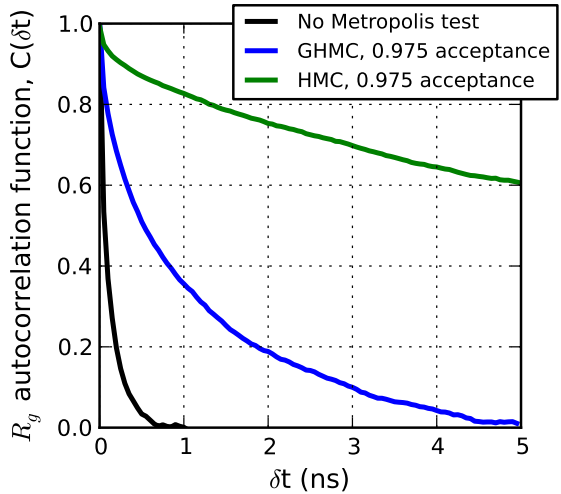}
\caption{\label{f:RgHMC} Comparison of the autocorrelation function for a polymer's radius of gyration obtained from simulations with no Metropolis test and with the GHMC and HMC algorithms.  Simulations were performed with a 10 fs time step and a friction constant of $\gamma$ = 0.01 ps$^{-1}$.   See methods section for full simulation details.
}
\end{figure}

The momentum adjustments required by these algorithms elicit concern that their implementation will diminish simulation mixing times or encumber interpretation of system dynamics \cite{Kennedy1996,Akhmatskaya2009,Sohl-Dickstein2012}.  The results of the present study demonstrate that this is indeed the case:  figure \ref{f:RgHMC} shows that, even with a small timestep and 
high acceptance rate, use of the HMC and GHMC methods for Langevin dynamics with a low friction constant result in a dramatic increase in the time required for system decorrelation.   

Recent results have demonstrated that an implementation of GHMC with no momentum flipping results in simulation data that closely reproduce the desired distribution,  though the algorithm cannot be proven to satisfy the balance condition that ensures proper sampling \cite{Akhmatskaya2009}.  As we demonstrate below, these results suggest a level of quantitative similarity for the average acceptance rates for two states with identical coordinates but opposite momenta.  Though this equivalence is not generally true, 
we propose a method that uses a simulation's immediately available information--that of the previous, current, and candidate states--to reduce the rate at which momenta are flipped.  This method 
 represents a simple modification to the GHMC algorithm that is generally applicable to integration schemes and provides a significant benefit to simulation mixing times.

\section{Theory}

\subsection{Traditional GHMC}
\label{s:tGHMC}

Let $\vec{x} \in \Omega$ denote some state of our system observed with probability 
\begin{equation}
P(\vec{x}) = \frac{1}{Z} \mbox{exp}
 \left[ -\beta H \left( \vec{x} \right) \right]
\label{e:HamiltonianDist}
\end{equation}
\begin{equation} 
Z = \int_{\Omega}  d\vec{x} \mbox{exp} 
\left[ -\beta H \left( \vec{x} \right) \right]
\label{e:PartitionFunction}
\end{equation}
where $\beta = \frac{1}{kT}$ is the inverse thermal energy.  Also note that $P\left(\vec{x}\right) =  P\left(\vec{\tilde{x}}\right)$, the tilde denotes momenta-reversal, and the Hamiltonian \emph{H} is invariant under such a transformation.  
We will assume a general simulation framework, where a candidate state $\vec{y}$ is generated from current state $\vec{x}$ 
with an associated selection probability 
$S \left( \vec{x} , \vec{y} \right)$.  
It is assumed that $\vec{y}$ is generated using an integration method that is
 irreducible (signifying that it can reach any state of the system from any other state), necessary to ensure ergodicity once Metropolization is included. 
 For the overall transition probability  $T \left( \vec{x} , \vec{y} \right)$,  the detailed balance condition imposes that 
\begin{equation}
P(\vec{x}) T(\vec{x} , \vec{y}) =
 P(\vec{\tilde{y}}) T(\vec{\tilde{y}} , \vec{\tilde{x}})
\label{e:DetailedBalance}
\end{equation}
The Metropolis criterion satisfies this condition:
\begin{equation}
  P_{\text{A}} \left( \vec{x} ,\vec{y} \right)= 
\mbox{min} \left(1,
\frac{S\left( \vec{\tilde{y}} , \vec{\tilde{x}} \right) }{S\left( \vec{x}, \vec{y} \right) }
\frac{P(\vec{\tilde{y}})}{P(\vec{x})}
\right)
\label{e:Metropolis}
\end{equation}

There are a number of methods appropriate for generating candidate state $\vec{y}$ in this framework, including Langevin  integrators \cite{Iii1984,Skeel2002,Paterlini1998,Bussi2007b,Sivak2012}, integration using the Andersen \cite{Andersen1980} or stochastic velocity rescaling thermostat \cite{Bussi2007}, and the traditional GHMC algorithm (leap-frog integration coupled with a mixing angle that introduces random velocity collisions) \cite{Horowitz1991} .  Simulations of this study generate candidate states with a single integration step using the velocity verlet with velocity randomization (VVVR) integrator \cite{Sivak2012}, equivalent to the Langevin integration scheme of Bussi and Parinello \cite{Bussi2007b}, or the stochastic velocity rescaling thermostat \cite{Bussi2007}.  It should be noted that the stochastic velocity rescaling thermostat appears to exhibit irreducible behavior \cite{Bussi2007}, but this has not been rigorously proven.   Selection probabilities for the stochastic velocity rescaling thermostat are given in appendix \ref{appendix:vrescale} and can be obtained for the VVVR integrator using the absolute path action derived in reference \cite{Sivak2012}.

We wish to demonstrate that, when coupled with either momentum reversal upon rejection or reversal upon acceptance, equation \ref{e:DetailedBalance} satisfies the balance condition 
$P\left( \vec{x} \right) = \int_{\Omega} d\vec{y} P(\vec{y}) S(\vec{y} , \vec{x}) P_{\text{A}}(\vec{y} , \vec{x})$.  This can be shown following a previously outlined proof \cite{Lelievre}, for which
we denote the average acceptance rate for transitions initiated from $\vec{x}$:
\begin{equation}
r(\vec{x}) = \int d\vec{y} S(\vec{x} , \vec{y}) P_{\text{A}}(\vec{x} , \vec{y})
\label{e:AcceptanceRate}
\end{equation}
If we begin in state $\vec{y}$ which we assume to have been drawn from the appropriate distribution with probability $P(\vec{y})$,
the next value in our chain, $\vec{x}$, is drawn from a probability $P^*( \vec{x})$ with two contributions:
 acceptance of the candidate $\vec{x}$ from initial state $\vec{y}$, and rejection of some candidate initiated from $\vec{y}=\vec{\tilde{x}}$.  Summing these contributions and using detailed balance:
\begin{eqnarray}
P^*\left( \vec{x} \right) & = & \int d\vec{y} P(\vec{y}) S(\vec{y} , \vec{x}) P_{\text{A}}(\vec{y} , \vec{x})
+ P(\vec{\tilde{x}}) \left[ 1- r(\vec{\tilde{x}})\right]
\nonumber \\ & = &
 \int d\vec{y} P(\vec{\tilde{x}}) S(\vec{\tilde{x}} , \vec{\tilde{y}}) P_{\text{A}}(\vec{\tilde{x}} , \vec{\tilde{y}})
+ P(\vec{\tilde{x}}) \left[ 1- r(\vec{\tilde{x}})\right]
\nonumber \\ & = &
P(\vec{\tilde{x}}) \left( r(\vec{\tilde{x}})+ 
\left[ 1- r(\vec{\tilde{x}})\right] \right) = P(\vec{x})
\label{e:BalanceProof}
\end{eqnarray}
The balance condition is also satisfied if the algorithm implements momentum flipping upon move \emph{acceptance} rather than rejection \cite{Nilmeier2011}.

Recent results have demonstrated that an implementation of GHMC with no such momentum flipping is capable of accurately reproducing certain statistical properties of molecular simulations  \cite{Akhmatskaya2009}.   This is a curious result, given that this algorithm cannot be shown to satisfy the balance condition.  We note, however, that if some configuration $\vec{x}$ and the associated $\vec{\tilde{x}}$ have identical acceptance rates, 
$r(\vec{x}) = r(\vec{\tilde{x}})$, then the balance condition is indeed satisfied without the need for momentum flipping:
Following the proof given in equation \ref{e:BalanceProof}, this can be shown assuming that  $r(\vec{x}) = r(\vec{\tilde{x}})$ and summing the appropriate contributions:
\begin{eqnarray}
P^*\left( \vec{x} \right) & = & \int d\vec{y} P(\vec{y}) S(\vec{y} , \vec{x}) P_{\text{A}}(\vec{y} , \vec{x})
+ P(\vec{x}) \left[ 1- r(\vec{x})\right]
\nonumber \\ & = &
P(\vec{\tilde{x}}) \left( r(\vec{\tilde{x}})+ 
\left[ 1- r(\vec{x})\right] \right) = P(\vec{x})
\label{e:BalanceProofAssume}
\end{eqnarray} 
Though we generally expect that $r(\vec{x}) \neq r(\vec{\tilde{x}})$, the aforementioned results  \cite{Akhmatskaya2009} suggest that $r(\vec{x})$ and $r(\vec{\tilde{x}})$ may have a level of quantitative similarity for some (perhaps easily met) set of conditions.
Here, we develop a moveset that expands on this concept to use information from both the candidate configuration  and the most recent configuration to reduce the probability of flipping momenta upon rejection while rigorously satisfying balance.  

\subsection{Reduced-flipping GHMC}

For the traditional GHMC framework, 
Sohl-Dickstein \cite{Sohl-Dickstein2012} has shown that, with proper accounting of the rates of inflow and outflow from $\vec{x}$,  the number of momentum flips upon candidate rejection can be reduced while still ensuring satisfaction of the balance condition.  This concept can be extended to algorithms that propagate coordinates and momenta in any fashion as long as the selection probability $S$ is well defined.

Our method, referred to as reduced-flipping GHMC, will select some candidate state ($\vec{z}$) from the current state ($\vec{y}$) according to the  probability $S\left( \vec{y} , \vec{z} \right)$.  The acceptance probability  for this move, $P_{\text{A}}(\vec{y},\vec{z})$,  obeys the standard Metropolis criterion given in equation \ref{e:Metropolis}.  Upon rejection, we will define probabilities for flipping ($\vec{y}\rightarrow \vec{\tilde{y}}$) and self ($\vec{y}\rightarrow \vec{y}$) transitions that depend on both the candidate $\vec{z}$ and the previous state $\vec{x}$.  These transitions are associated with probabilities 
 $P_{\text{S}}(\vec{y} | \vec{x},\vec{z}) $ for self-transitions and $P_{\text{F}}(\vec{y},\vec{\tilde{y}} | \vec{x},\vec{z}) $ for momentum flipping transitions.  Thus, this algorithm proceeds as follows
\begin{enumerate}
\item Select a candidate state $\vec{z}$ from current state $\vec{y}$ according to the probability $S\left( \vec{y} , \vec{z} \right)$. 
\item Generate a uniform random number, $\xi \in \left[0,1\right)$ and perform the transition $\left( \vec{y} \rightarrow \vec{y}_2 \right)$, where
 \begin{equation}
\vec{y}_2 = \left\{ 
\begin{array} {l l }
\vec{z} \qquad \mbox{if } \xi < P_{\text{A}}(\vec{y},\vec{z})
\nonumber \\
\vec{y} \qquad \mbox{if } P_{\text{A}}(\vec{y},\vec{z}) \leq \xi < \left(P_{\text{A}}(\vec{y},\vec{z})+P_{\text{S}}(\vec{y} | \vec{x},\vec{z})\right)
\nonumber \\
\vec{\tilde{y}} \qquad \mbox{else}
\end{array}
\right. 
\end{equation}
\end{enumerate}
where $P_{\text{A}}(\vec{y},\vec{z})+P_{\text{S}}(\vec{y} | \vec{x},\vec{z})+ P_{\text{F}}(\vec{y},\vec{\tilde{y}} | \vec{x},\vec{z}) = 1$.

The traditional GHMC method described in section \ref{s:tGHMC}, henceforth referred to as the automatic-flipping algorithm, always flips momenta upon candidate rejection and corresponds to enforcing the constraints  $P_{\text{S}}(\vec{y} | \vec{x},\vec{z}) = 0 $ and 
 $P_{\text{F}}(\vec{y},\vec{\tilde{y}} | \vec{x},\vec{z}) = 1- P_{\text{A}}(\vec{y},\vec{z})$.   For the reduced-flipping method, we wish to remove these constraints and minimize $P_{\text{F}}$ while imposing:
 \begin{eqnarray}
 &&
 P(\vec{x}) S(\vec{x},\vec{y}) P_{\text{A}} (\vec{x},\vec{y}) S(\vec{y},\vec{z}) P_{\text{F}}(\vec{y},\vec{\tilde{y}} | \vec{x},\vec{z}) 
  \nonumber \\ 
 && \quad 
 + 
 P(\vec{\tilde{z}}) S(\vec{\tilde{z}},\vec{\tilde{y}} )  P_{\text{A}} (\vec{\tilde{z}},\vec{\tilde{y}} )  S(\vec{y},\vec{x})
 P_{\text{S}}(\vec{\tilde{y}} | \vec{\tilde{z}},\vec{\tilde{x}})
 \nonumber \\ 
 && \quad =
 P(\vec{x}) S(\vec{x},\vec{y})  P_{\text{A}}  (\vec{x},\vec{y})  S(\vec{y},\vec{z}) \left(  
 1- P_{\text{A}}(\vec{y},\vec{z})
 \right) 
 \label{e:PRelation}
 \end{eqnarray}
 noting that 
 \begin{equation}
  P_{\text{F}}(\vec{y},\vec{\tilde{y}} | \vec{x},\vec{z}) \in
  \left[ 0,  1- P_{\text{A}}(\vec{y},\vec{z}) \right]
 \end{equation}
 \begin{equation}
 P_{\text{S}}(\vec{\tilde{y}} | \vec{z},\vec{x}) 
 \in \left[ 0, 1- P_{\text{A}}(\vec{\tilde{y}},\vec{\tilde{x}})\right]
 \end{equation}
 Thus, equation \ref{e:PRelation} considers the proposed forward transition $\vec{y}\rightarrow\vec{z}$ given $\vec{x}$ and
 the proposed reverse transition $\vec{\tilde{y}}\rightarrow \vec{\tilde{x}} $ given $ \vec{\tilde{z}}$.  By considering the overall rejection rates,  we can ensure that the number of transitions to state $\vec{\tilde{y}}$ by the reduced-flipping algorithm (given by the left hand side of equation \ref{e:PRelation}) remains equivalent to the overall rejection rate of the proposed forward move (the right hand side of equation \ref{e:PRelation}) while  
 minimizing the number of momentum flips resulting from each proposed transition.   This method breaks the detailed balance condition for self-transitions, corresponding to the modified condition:
\begin{equation}
P(\vec{x}) T(\vec{x} , \vec{y}) =
 P(\vec{\tilde{y}}) T(\vec{\tilde{y}} , \vec{\tilde{x}}) ,  \vec{x} \neq \vec{y}
\label{e:DetailedBalanceModified}
\end{equation}

This reduced-flipping method applies only to the case in which the previous transition resulted from an \emph{accepted} move.  
If the previous transition resulted from a rejected move, we will use 
the standard GHMC criteria (automatic momentum flipping).   This limitation is discussed in more detail below.  

The probabilities given the above constraints are easily defined:

\begin{widetext}
%
%
 \begin{equation}
P_{\text{S}}(\vec{y} | \vec{x},\vec{z}) = \left\{ 
\begin{array} {l l } 
\mbox{min} \left(
1 - P_{\text{A}}(\vec{y},\vec{z}) ,
\frac{
P(\vec{\tilde{z}}) S(\vec{\tilde{z}},\vec{\tilde{y}})  P_{\text{A}}  (\vec{\tilde{z}},\vec{\tilde{y}} )  S(\vec{\tilde{y}},\vec{\tilde{x}})
}
{
P(\vec{x}) S(\vec{x},\vec{y})  P_{\text{A}} (\vec{x},\vec{y})   S(\vec{y},\vec{z} ) 
}
\left( 1 -  P_{\text{A}}(\vec{\tilde{y}},\vec{\tilde{x}})  \right) 
\right)
\quad
 \mbox{if }  \vec{x} \rightarrow \vec{y}
  \mbox{ was an}  \\ 
  \qquad \qquad \qquad \qquad \qquad   \qquad \qquad \qquad \qquad \qquad \qquad \qquad \qquad 
  \quad 
  \mbox{  accepted transition} 
  \\ 
0 \qquad \qquad \qquad \qquad \qquad   \qquad \qquad \qquad \qquad \qquad \qquad \qquad \quad 
 \mbox{else } 
\end{array}
\right.
\label{e:SelfProb}
\end{equation}
 
  \begin{equation}
  \begin{array} {l l } 
P_{\text{F}}(\vec{y} , \vec{\tilde{y}} | \vec{x},\vec{z}) = 1- P_{\text{A}}(\vec{y},\vec{z}) -
P_{\text{S}}(\vec{y} | \vec{x},\vec{z}) & \qquad \qquad \qquad
 \qquad \qquad \qquad  \qquad \qquad \qquad  \qquad \qquad \qquad \qquad
\end{array}
\label{e:FlipProb}
\end{equation} 
\end{widetext}
 Equations \ref{e:SelfProb} and \ref{e:FlipProb}, which define the probabilities of self- and flipping-transitions for the reduced-flipping GHMC method, are the central result of this manuscript.

 Once again, we wish to prove satisfaction of the balance condition by equating $P^* (\vec{y})$, the probability of observing $\vec{y}$ within our simulation, to the stationary distribution $P (\vec{y})$.  $P^* (\vec{y})$ receives contributions from all accepted transitions to $\vec{y}$ 
 and from rejected candidates from the initial states $\vec{y}$ and $\vec{\tilde{y}}$.  

\begin{widetext}
\begin{eqnarray}
 P^* (\vec{y}) & = &  \int_{\Omega} d \vec{x} P(\vec{x}) S (\vec{x},\vec{y}) P_{\text{A}}  (\vec{x},\vec{y})
 +  \int_{\Omega} d \vec{x}  \int_{\Omega} d \vec{z} \left[  
 P(\vec{x}) S(\vec{x},\vec{y}) P_{\text{A}} (\vec{x},\vec{y})   S(\vec{y},\vec{z}) 
 P_{\text{S}}(\vec{y} | \vec{x},\vec{z}) 
 \right.  \nonumber \\ && \qquad \qquad \qquad  \qquad \qquad \qquad \qquad  \qquad \left.
 +
 P(\vec{x}) S(\vec{x},\vec{\tilde{y}})  P_{\text{A}} (\vec{x},\vec{\tilde{y}}) S(\vec{\tilde{y}},\vec{z}) 
 P_{\text{F}}(\vec{\tilde{y}},\vec{y} | \vec{x},\vec{z}) 
   \right]  
   \nonumber \\ 
   & = &  \int_{\Omega} d \vec{x} 
   P(\vec{\tilde{y}}) S(\vec{\tilde{y}} , \vec{\tilde{x}})  P_{\text{A}}  (\vec{\tilde{y}} , \vec{\tilde{x}})
 +  \int_{\Omega} d \vec{x}  \int_{\Omega} d \vec{z} \left[  
 P(\vec{x}) S(\vec{x},\vec{y}) P_{\text{A}} (\vec{x},\vec{y}) S(\vec{y},\vec{z}) 
 P_{\text{S}}(\vec{y} | \vec{x},\vec{z}) 
 \right.  \nonumber \\ && \qquad \qquad \qquad  \qquad \qquad \qquad \qquad  \qquad \left. 
 +
 P(\vec{\tilde{z}}) S(\vec{\tilde{z}},\vec{\tilde{y}})  P_{\text{A}}  (\vec{\tilde{z}},\vec{\tilde{y}})  S(\vec{\tilde{y}},\vec{\tilde{x}}) 
 P_{\text{F}}(\vec{\tilde{y}},\vec{y} | \vec{\tilde{z}},\vec{\tilde{x}}) 
   \right]  
   \nonumber \\ 
    & = &  
   P(\vec{\tilde{y}}) r(\vec{\tilde{y}})  
  +  \int_{\Omega} d \vec{x}  \int_{\Omega} d \vec{z} 
 P(\vec{x}) S(\vec{x},\vec{y}) P_{\text{A}} (\vec{x},\vec{y})   S(\vec{y},\vec{z})  (1-P_{\text{A}}  (\vec{y},\vec{z}))
 \nonumber \\ & = & 
   P(\vec{\tilde{y}}) r(\vec{\tilde{y}})  
  +  \int_{\Omega} d \vec{x}  \int_{\Omega} d \vec{z} 
 P(\vec{\tilde{y}}) S(\vec{\tilde{y}},\vec{\tilde{x}})  P_{\text{A}} (\vec{\tilde{y}},\vec{\tilde{x}})  S(\vec{y},\vec{z})  (1-P_{\text{A}}  (\vec{y},\vec{z}))
  \nonumber \\ & = & 
   P(\vec{\tilde{y}}) r(\vec{\tilde{y}})  
  +   P(\vec{\tilde{y}})  \int_{\Omega} d \vec{z} 
   S(\vec{y},\vec{z})  (1-P_{\text{A}}  (\vec{y},\vec{z}))
    \nonumber \\ & = & 
   P(\vec{\tilde{y}}) r(\vec{\tilde{y}})  
  +   P(\vec{\tilde{y}}) \left(1-r(\vec{\tilde{y}})  \right) = P(\vec{y})
 \label{e:Mybalance}
 \end{eqnarray} 
 \end{widetext}
 
As is demonstrated by equation \ref{e:FlipProb}, the reduced-flipping method reverts to the automatic-flipping algorithm if the  previous transition was a rejected move.  
This is because the decrease in momentum-flipping transitions relies on the detailed balance condition given in equation \ref{e:DetailedBalanceModified}, which cannot be applied to transitions resulting from rejected moves.  
It is possible to remove this limitation by establishing self- and flipping-transition probabilities that depend on the \emph{path} of rejections (and, thus, every rejected candidate state) taken following the last accepted transition.  This modification adds significant complexity to the algorithm (the calculation of such probabilities scale as $O(N!)$ for \emph{N} serial rejections) and is unlikely to add significant benefit to simulation mixing times.

\section{Methods}
All simulations were performed using a modified version of GROMACS-3.3.1 \cite{berendsen95,Lindahl2001} and the MARTINI 2.1 forcefield \cite{Marrink2007,Marrink2008}.  Results were collected on an eight-residue repeat of polyglutamine immersed in water at 400K using either the VVVR integrator \cite{Sivak2012} or verlet integration with stochastic velocity rescaling \cite{Bussi2007}.  Simulations using VVVR integration were performed over a range of values for the friction coefficient, $\gamma \in \left\{0.01, 1.0, 91  \right\}$ ps$^{-1}$.  
 Simulations were performed using nonbonded interactions that are switched at 0.9 nm and cut off at 1.2 nm in conjunction with a dispersive tail correction \cite{Shirts2007}.

As an indication of mixing times for the algorithms presented in this work, we study the autocorrelation function for the radius of gyration, $R_g$, for the polyglutamine octamer:
\begin{equation}
C \left(\delta t \right) = \frac{\left \langle R_g(t)R_g(t+\delta t)\right \rangle}{\left \langle R_g^2 \right \rangle}
\end{equation}
We also analyze the integrated autocorrelation function, $A_{R_g} = \int_0^{\infty} C \left(  \delta t \right)  d \delta t$.  Samples were collected every 10 ps and 
 statistical uncertainties for integrated autocorrelation functions were estimated using bootstrapping analysis.

 \begin{figure}
 \centering
\includegraphics[width=8.0 cm]{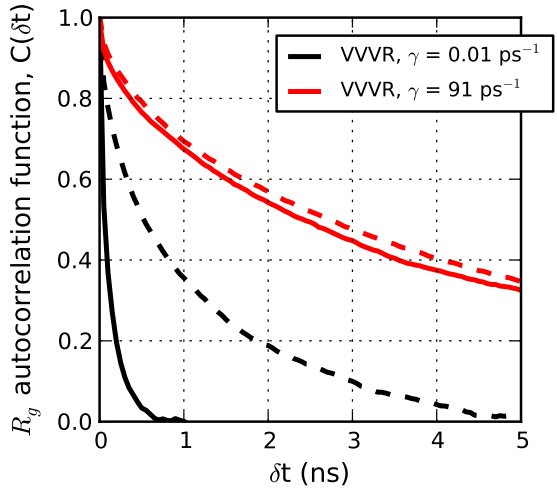}
\caption{\label{f:Gamma}  
C(t), the radius of gyration autocorrelation function, calculated using the VVVR integrator at both high ( $\gamma$ = 91 ps$^{-1}$)
and low ( $\gamma$ = 0.01 ps$^{-1}$) viscosity.  
All simulations were performed using a 10 fs time step and resulted in at least a 97\% acceptance rate.  Solid curves were obtained with no Metropolis test while dashed curves were obtained using the traditional GHMC automatic-flipping algorithm.  As can be seen, use of this automatic-flipping algorithm significantly extends the time required for system decorrelation for simulations performed with a small friction coefficient, while this effect is very small for simulations performed with a high friction constant. 
}
\end{figure}

 \begin{figure}
 \centering
\includegraphics[width=8.0 cm]{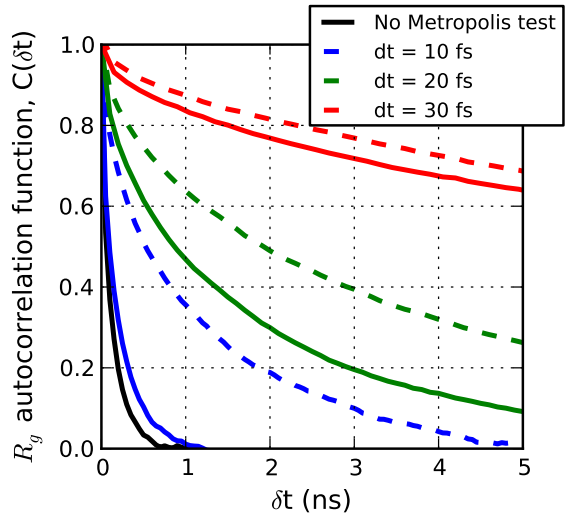}
\caption{\label{f:AFSF}C(t), the radius of gyration autocorrelation function, using a simulation performed with no Metropolis test as a standard comparison.  Simulations were performed using the VVVR integrator with $\gamma$ = 0.01 ps$^{-1}$. Dashed lines correspond to results obtained using the traditional, automatic-flipping algorithm while solid lines correspond to those obtained with the reduced-flipping algorithm. 
}
\end{figure}

\begin{table*}
	\centering
	\caption[]{
	 Comparison of the integrated autocorrelation function for simulations performed with no Metropolis test, traditional GHMC with automatic momentum flipping, and reduced-flipping GHMC.
	}
	\small
	\begin{tabular}{|c|c|c|c|}
		\hline 
		\diaghead{\theadfont LotsandLotsofmorestuff}{Algorithm}{ Timestep (fs)}
		& 10 & 20 & 30 \\ \hline
		\multicolumn{4}{|c|}{VVVR, $\gamma$ = 91 ps$^{-1}$} \\ \hline
		$A_{R_g}$(ns), no Metropolis test &   4.61 $\pm$  0.08         &    --         & --  \\ 
		$A_{R_g}$(ns), automatic-flipping  &  5.41 $\pm$ 0.12             & 6.10 $\pm$ 0.05           & 10.23 $\pm$ 0.38   \\ 
		$A_{R_g}$(ns), reduced-flipping  & 4.99  $\pm$ 0.08             &  5.50 $\pm$ 0.04         &   10.08 $\pm$ 0.51  \\
		 \hline	
		\multicolumn{4}{|c|}{VVVR, $\gamma$ = 1.0 ps$^{-1}$} \\ \hline
		$A_{R_g}$(ns), no Metropolis test & 0.14  $\pm$ 0.01                  &  -- & --  \\ 
		$A_{R_g}$(ns), automatic-flipping &  0.52 $\pm$  0.01               &  1.52 $\pm$ 0.06                  & 11.50 $\pm$ 0.30 \\ 
		$A_{R_g}$(ns), reduced-flipping & 0.22  $\pm$  0.01               &   1.04 $\pm$0.05                 &   7.11 $\pm$  0.17 \\
		 \hline	
		\multicolumn{4}{|c|}{VVVR, $\gamma$ = 0.01 ps$^{-1}$} \\ \hline
		$A_{R_g}$(ns), no Metropolis test & 0.12  $\pm$ 0.01                  &  -- & --  \\ 
		$A_{R_g}$(ns), automatic-flipping & 1.04 $\pm$ 0.01                 &  3.46 $\pm$ 0.09                  & 23.04  $\pm$ 0.31 \\ 
		$A_{R_g}$(ns), reduced-flipping &  0.16 $\pm$  0.01               &   2.00 $\pm$ 0.04                & 17.96   $\pm$ 0.42 \\
		 \hline	
		\multicolumn{4}{|c|}{Stochastic velocity rescaling} \\ \hline
		$A_{R_g}$(ns), no Metropolis test & 0.07 $\pm$  0.01  &  -- & --  \\ 
		$A_{R_g}$(ns), automatic-flipping &  1.62 $\pm$ 0.02                &  5.21 $\pm$ 0.09                      &  28.11  $\pm$ 0.19      \\ 
		$A_{R_g}$(ns), reduced-flipping &  0.22 $\pm$ 0.02                &   2.07 $\pm$  0.04                    & 20.93  $\pm$ 0.16   \\
		 \hline		 		 
	\end{tabular}
	\label{tab:C}
\end{table*}

\section{Results}

Figure \ref{f:Gamma} compares the autocorrelation functions for a polymer radius of gyration calculated from simulations using the traditional GHMC (automatic-flipping) algorithm and from those with no Metropolis test.  These simulations were performed with a 10 fs time step, resulting in an acceptance rate of at least 97\% for all GHMC simulations. As can be seen, the effect of the automatic-flipping algorithm is insignificant for simulations performed with the VVVR integrator and a friction constant of $\gamma$ = 91 ps$^{-1}$.  The algorithm is shown to significantly extend the time required for system decorrelation, however, as the the friction constant is decreased.  
This effect is most pronounced for results obtained using the stochastic velocity rescaling thermostat.  This is not surprising, given that this integration scheme can be derived starting with Langevin integration and minimizing the thermostat's disturbance on the Hamiltonian trajectory \cite{Bussi2008}.  Table \ref{tab:C} gives the integrated autocorrelation function for these results.  Again, for the most pronounced case using the stochastic velocity rescaling thermostat, the $A_{R_g}$ value increases over an order of magnitude, from 0.07 $\pm$ .01 ns for a simulation containing no Metropolis test to 1.62 $\pm$ 0.02 ns for tradtional GHMC.

For simulations performed using the VVVR integrator with low viscosity ($\gamma$ = 0.01 ps$^{-1}$), figure \ref{f:AFSF} displays autocorrelation functions for a polymer radius of gyration calculated using the automatic-flipping and reduced-flipping algorithms.  The effect of the GHMC method on system dynamics is substantially mitigated by use of the reduced-flipping algorithm.  For simulations performed with the stochastic velocity rescaling thermostat, $A_{R_g} =$ 0.22 $\pm$ 0.02 ns for the reduced-flipping algorithm.  This is very close to the value obtained with no Metropolis test ($A_{R_g}$ = 0.07 $\pm$ .01 ns), in contrast to the increase by an order of magnitude observed for the automatic-flipping algorithm ($A_{R_g}$ = 1.62 $\pm$ 0.02 ns).

The  benefit of reduced-flipping GHMC becomes negligible at large time steps and small acceptance rates (for dt = 30 fs, these GHMC simulations resulted in  acceptance rates $\approx$ 40\%).  Unsurprisingly, the reduced-flipping algorithm also demonstrates no major benefit for simulations performed at high viscosity (see table \ref{tab:C} for a comparison of automatic-flipping and reduced-flipping algorithms over all integration parameters).

\section{Conclusions}

The introduction of a Metropolis criterion into a molecular dynamics simulation traditionally requires an undesirable 
transformation of the system momenta, either via complete randomization (HMC) or momentum flipping upon move rejection (GHMC). This manuscript presents results obtained using a coarse-grained forcefield simulation of a polyglutamine octamer and demonstrates that concerns over the effect of these algorithms on simulation mixing times may be well justified, depending on the particular method of integration. 
The two main conclusions of this paper are that (1) for certain sets of conditions, transformations of system momenta associated with HMC/GHMC algorithms significantly affect simulation mixing times, and (2) our reduced-flipping algorithm presents a simple alternative to the traditional GHMC algorithm that significantly reduces this effect.  These results have been demonstrated for both the VVVR integrator, equivalent to Langevin dynamics \cite{Sivak2012}, and the stochastic velocity rescaling thermostat, which can be derived as a form Langevin integration by minimizing the thermostat's disturbance on the Hamiltonian trajectory \cite{Bussi2008}.

The differences between results obtained with no Metropolis test, with traditional automatic-flipping GHMC, and with the reduced-flipping GHMC are insignificant for simulations performed with a high friction coefficient but become quite pronounced for integration methods corresponding to low friction or velocity collision rate.  Surprisingly, these effects are significant even at small time steps associated with high acceptance rates.  The reduced-flipping algorithm presented in this work, a simple modification to the traditional GHMC method that rigorously satisfies the balance condition, is capable of significantly reducing the effect that these automatic-flipping algorithms have on molecular kinetics and simulation mixing times. 
 
\section{Acknowledgements}
We are very grateful to John Chodera for thoughtful discussion of this manuscript.  Computer resources were provided by the National Science Foundation award 0960306. 

\section{Appendix}
\appendix

\section{The selection probability for integration steps using a stochastic velocity rescaling thermostat}
\label{appendix:vrescale}
Thermostatting via stochastic velocity rescaling \cite{Bussi2007} is similar to the Berendsen thermostat \cite{berendsen1984} with an additional stochastic term that ensures proper canonical sampling. 
The positions and velocities are propagated using standard integration methods.  At iteration \emph{i}, the kinetic energy ($E_{\text{k}}^i$) is updated:
\begin{equation}
E_{\text{k}}^{i*} = E_{\text{k}}^i + (E_{\text{k}}^{\text{ref}} - E_{\text{k}}^i)\frac{\Delta t}{\tau} + 2 \sqrt{\frac{E_{\text{k}}^iE_{\text{k}}^{\text{ref}}}{N_f \tau}}\xi
\label{e:Vrescale}
\end{equation}
where $E_{\text{k}}^{\text{ref}} = \frac{N_f}{2\beta}$ is the average kinetic energy given the number of degrees of freedom ($N_f$) and the inverse thermal energy ($\beta$),  the parameter $\tau$ defines the relaxation timescale, and $\xi$ is a Gaussian random variable of mean 0 and standard deviation 1. 
To achieve this change in kinetic energy, all velocities are scaled by the factor $\sqrt{\frac{E_{\text{k}}^{i*}}{E_{\text{k}}^i}}$. 

We can calculate the Gaussian random variable $\tilde{\xi}$ necessary for the reverse transition: 
\begin{equation}
\tilde{\xi} = E_{\text{k}}^i-E_{\text{k}}^{i*} - \frac{\left( E_{\text{k}}^{\text{ref}}-E_{\text{k}}^{i*} \right)\sqrt{N_f}}{2\sqrt{E_{\text{k}}^{i*}E_{\text{k}}^{\text{ref}}\Delta t \tau}}
\label{e:VRreverse}
\end{equation}
Note that we must include the Jacobian factor accounting for the coordinate transformation from $\xi$ (a single random variable applied to the total kinetic energy) to the Cartesian space of the $N_f$ momentum degrees of freedom. 
The ratio of selection probabilities for a transition  $\left(\vec{x}  \rightarrow \vec{y} \right)$ is:
\begin{equation}
\frac{S\left(\vec{\tilde{y}}  \rightarrow \vec{\tilde{x}} \right)}{S\left(\vec{x}  \rightarrow \vec{y} \right)}
 = 
\left(
\frac{E_{\text{k}}^{i*} }{E_{\text{k}}^{i} }
\right)^{\left(N_f-3\right)/2}
\mbox{exp} \left[  0.5 \left(
 \xi^2-\tilde{\xi}^2 \right) 
\right] 
\label{e:VrescaleSelection}
\end{equation}

\bibliographystyle{phjcp}
\bibliography{GHMC}

\end{document}